\documentclass{iopjournal}
\usepackage{soul}
\usepackage[sorting=none]{biblatex}
\begin{document}
\newcommand{\chasemac}[1]{\textcolor{blue}{CCM: #1}}
\newcommand{\nk}[1]{\textcolor{purple}{NK: #1}}
\newcommand{\SC}[1]{\textcolor{red}{SC: #1}}
\articletype{Article type} %

\title{The mechanical latching memory of an adhesive tape}

\author{Sebanti Chattopadhyay$^{1,*}$\orcid{0000-0000-0000-0000}, Carys Chase-Mayoral$^{1,2}$\orcid{0000-0000-0000-0000} and Nathan Keim$^{1}$\orcid{0000-0000-0000-0000}}

\affil{$^1$Department of Physics, The Pennsylvania State University, University Park, PA 16802, USA}

\affil{$^2$Department of Physics, Dickinson College, Carlisle, PA 17013, USA}

\affil{$^*$Author to whom any correspondence should be addressed.}

\email{sebanti@psu.edu}


\begin{abstract}
The storage and retrieval of mechanical imprints from past perturbations is a central theme in soft matter physics. Here we study this effect in the partial peeling of an ordinary adhesive tape, which leaves a line of strong adhesion at the stopping point. We show how this behavior can be used to mechanically store and retrieve the amplitudes of successive peeling cycles. This multiple-memory behavior resembles the well-known return-point memory found in many systems with hysteresis, but crucially the driving here is rectified: peeling is unidirectional, where each cycle begins and ends with the tape flat on the substrate. This condition means that the tape demonstrates a distinct principle for multiple memories. By considering another mechanism that was recently proposed, we establish ``latching'' as a generic principle for memories formed under rectified driving, with multiple physical realizations. We show separately that tape can be tuned to erase memories partially and also demonstrate the function of tape as a mechanical computing device that extracts features from input sequences and compares successive values. 

\end{abstract}

\section{Introduction}
\begin{figure}
 \centering
        \includegraphics[width=0.9\textwidth]{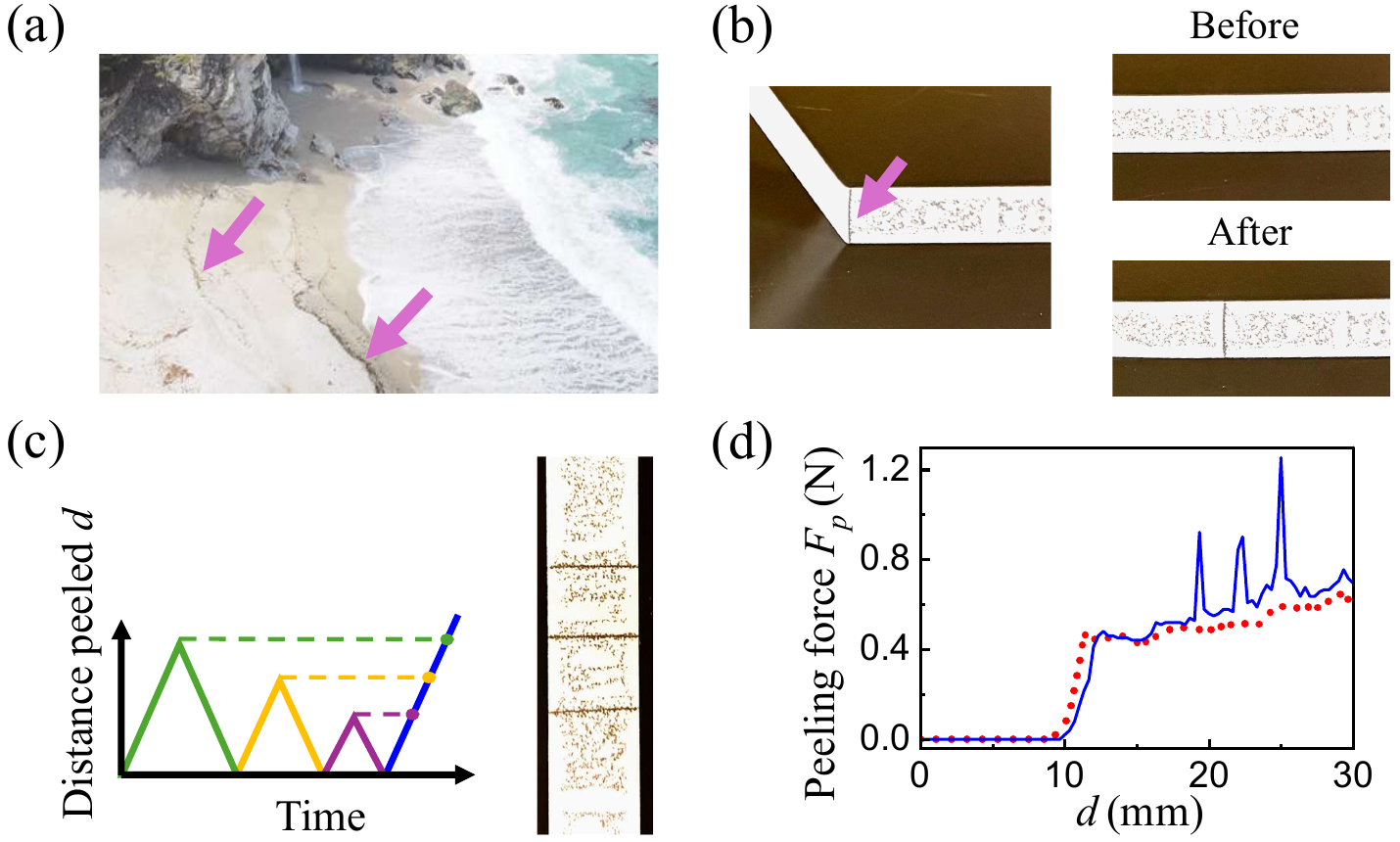}
 \caption{\textbf{Memories of turning points.} (a) Strandlines (indicated by arrows) on the beach at Julia Pfeiffer Burns State Park, California, USA. 
 (b) An incompletely peeled adhesive tape, of width 1.3~cm, retains a line of strong adhesion at the turning point where peeling stopped (indicated by the arrow), resembling a strandline in (a). ``Before'' and ``after'' pictures show the tape laid flat. Dark speckles indicate sparse contact with the substrate, corresponding to weaker adhesion. 
 (c) Schematic of the protocol used to encode multiple turning points by peeling to successively smaller distances. After encoding we apply readout (final, blue segment), peeling past all stored values. Side panel shows the tape with visible imprints of the three encoded points. 
 (d) Peeling force $F_p$ during readout with encoded distances at $d= 20, 22.5,$ and $25$~mm. Distinct peaks in $F_p$ correspond to the encoded $d$ values. Dotted line is from the same tape with no memories.}
\label{fig1}
\end{figure}

Among the beach's many fascinations are the countless shells, stones, and debris that wash ashore. These pieces have a captivating variety, but often their placement has a striking order: instead of being strewn about at random by the waves, they form sharp lines that run along the sand (Fig.~1(a)). These strandlines mark where past waves turned back---the beach carries a partial history of the tides \cite{Gheskiere2006Apr}. A simplified version of strandlines is an adhesive tape. Peeling the tape to a distance $d$ and laying it back down leaves a line of strong adhesion at $d$, as shown in Fig.~1(b). Repeating cycles of this with successively smaller $d$ stores additional memories (Fig.~1(c)). The visible imprint of each stored memory leads to a corresponding spike in the peeling force (Fig.~1(d))---both writing and reading can be purely mechanical. In this paper, we examine how multiple memories are written, read, and erased in an adhesive tape, including how memory strength and erasure can be tuned. The features of this memory allow it to perform simple information processing functions, like one-back comparison tasks \cite{Szmalec2009May, Jaeggi2010May}. We propose that the tape illustrates a generic principle for storing sequences that is distinct from other known examples of material memory. 

The unifying feature of the strandlines and the tape is the storage of turning points---a basic way to extract one or more features from the history of a scalar drive. 
The same concept is put to practical use in the single-dial combination lock. To unlock, the user turns the dial back and forth by decreasing amounts, reversing at each successive value in the combination.
The lock's ability to store the sequence of turning points is a consequence of return-point memory, a generic property of ferromagnets~\cite{Adair1983Apr}, sandstone, and many other systems that can be modeled as ensembles of hysteretic elements~\cite{Keim2019Jul,Paulsen2025Mar}. Return-point memory also means that as long as the driving is bounded by two turning points, the system must have the same state each time a turning point is revisited~\cite{Adair1983Apr,Sethna1993May}. In light of this definition, we see that while the driving sequence in Fig.~1(c) stores multiple memories in tape, it will fail in every system with return-point memory. The reason is that the driving in this case is rectified: peeling cannot be negative, and after the first (largest) value is stored, the second and third cycles revisit the turning point at zero, which will reset any system with return-point memory and preclude multiple memories~\cite{Lindeman2025Jan}. The rectified polarity of driving means that a different paradigm is needed for tape.

Despite the many positive-definite quantities in nature, relatively few studies have highlighted how multiple memory formation can depend on the polarity of driving. Some disordered systems are known to be agnostic to this polarity, but they are unlike tape in that reliably forming $N$ memories tends to require more than $N$ cycles of driving. Chattopadhyay and Majumdar~\cite{Chattopadhyay2022Jun} found that granular gels store multiple turning points of shear strain encoded in a decreasing order (as in Fig.~1c), leading to sharp changes in stress upon subsequent shearing (as in Fig.~1d). However, the smaller amplitude must be repeated several times for retention of the memory. Dilute suspensions of non-adhesive particles~\cite{Paulsen2014Aug, Padamata2025Oct} show another generic behavior called multiple transient memories, in which multiple memories form with repeated training, but cannot persist after a large number of cycles unless memory formation competes with noise~\cite{Keim2019Jul,Paulsen2014Aug,Keim2011Jun}. In this case, encoding multiple memories is insensitive to the ordering of the amplitudes. More recently, Lindeman et al.\ identified a ``latching'' mechanism involving pairs of coupled bistable elements called hysterons, which can store $N$ amplitudes from $N$ cycles of rectified driving---similar to the immediacy of return-point memory. This mechanism is likely to be rare in disordered systems, but has been demonstrated experimentally with artificial hysterons that are fine-tuned for the behavior~\cite{Paulsen2024Sep}. In systems with hysteretic elements that are not specially designed, it is unclear whether memories of rectified driving could ever be nearly as strong as memories of non-rectified driving, because the latter are stored by the same elements' robust return-point behavior~\cite{Lindeman2025Jan}.

Our experiments, in this work, with pressure-sensitive adhesive (PSA) tapes offer an unusually simple model system for studying memories under rectified driving. Memory effects are large (Fig.~1(d)) and arise from a clear physical mechanism: due to the bending stiffness of the tape, the region just ahead of the peeling front is pushed into the substrate, strengthening the adhesive bond~\cite{Kaelble1960Mar, Bartlett2023Aug}. This clarity allows us to discern simple rules for multiple memories. We also use the unique physics of PSA tape to vary the memory strength and introduce partial erasure, aspects that are rarely tunable in other memory-forming systems. Studying tape lets us reformulate the latching behavior identified by Lindeman et al.~\cite{Lindeman2025Jan}, without the finely-tuned pairs of bistable elements, yielding a more generic principle for studying and designing non-equilibrium matter that forms memories.

\section{Results}
Figures~1(b) and (c) illustrate the encoding of multiple memories in the tape by peeling to successively smaller distances. To read out the encoded values, we peel the tape to a distance larger than all training distances (Fig.~1(c)) while measuring the upward component of force $F_p$. All peeled distances are represented by the vertical displacement $d$ of the translation stage that peels and lays down the tape (see Materials and Methods). The displacement $d$ of the stage is approximately equal to the horizontal displacement of the peeling front. Fig.~1(d) shows $F_p$ for a tape with encoded distances $d = 20, 22.5$ and $25$~mm. At each encoded $d$ value, there is a spike that roughly doubles the peeling force. In contrast, the same tape without memories shows a nearly monotonic $F_p$ with no such spikes (dotted curve in Fig.~1(d)).  

\begin{figure}
 \centering
        \includegraphics[width=1\textwidth]{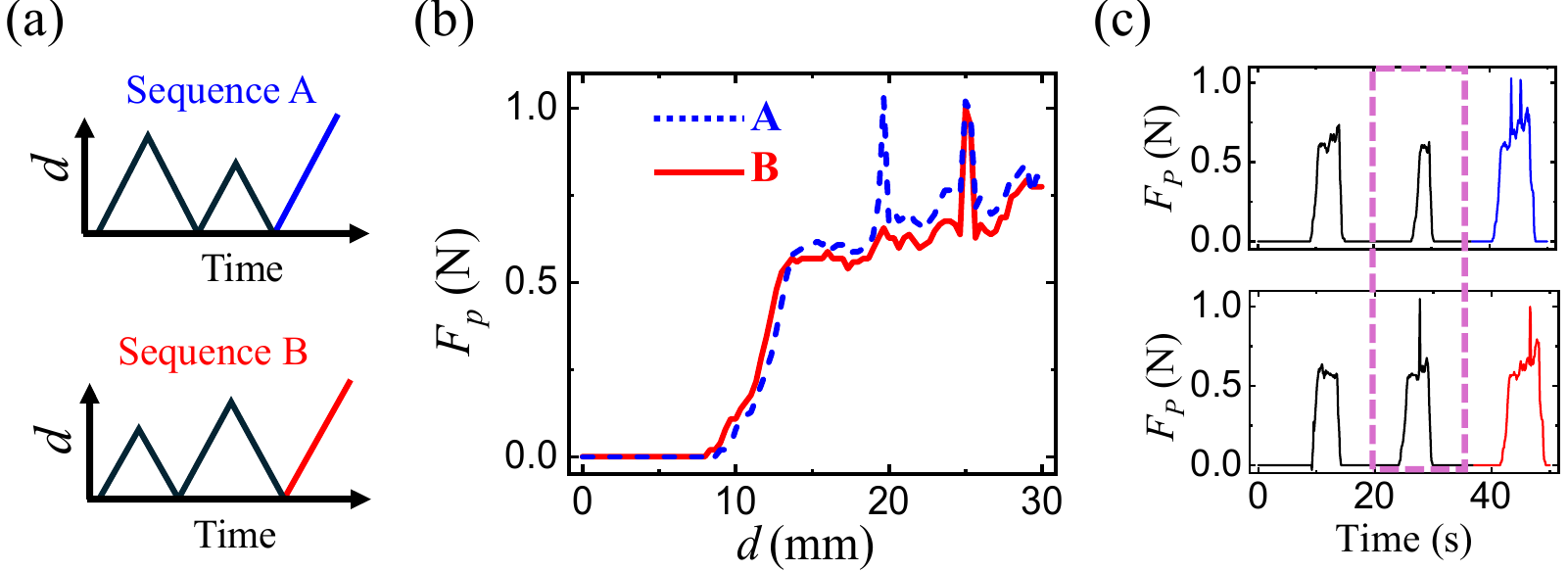}
 \caption{\textbf{Conditions for multiple memory retention.} 
(a) Protocols for encoding and readout of two memories at $d = 20~mm$ and $d = 25~mm$ in decreasing (Sequence A) and increasing (Sequence B) order. 
(b) Peeling force $F_p$ during readouts for both sequences. Both memories are present only if written in decreasing order.
(c) Peeling force during the entirety of each protocol. Pink box highlights the second write cycle. In Sequence B (bottom), the larger amplitude reads and erases the smaller value.}
\label{fig2}
\end{figure}

\subsection{Nested Memories}

Retention of multiple memories by a system may or may not depend on the ordering of the amplitudes during encoding. In previous studies, this question helped to distinguish among known memory behaviors~\cite{Keim2019Jul}. To study this in the tape, we compare two sequences A and B, where a large and a small distance ($d=20$~mm and 25~mm) are encoded in decreasing and increasing order, respectively (Fig.~2(a)). In Fig.~2(b), the spikes in the readout of Sequence A indicate retention of both memories, while Sequence B retains only the larger distance. We conclude that encoding the larger distance erased the smaller one, i.e. the signature of the memory became indistinct from other variations in the peeling force. Such an erasure behavior is typical in magnets and other systems with return-point memory~\cite{Keim2019Jul, Paulsen2025Mar}. In Fig.~2(c), we show $F_p$ during the entire process of encoding and readout for both sequences in Fig.~2(a). We see that the process of encoding the larger input in Sequence B must also read out the preceding smaller input, thus erasing its memory.  

\subsection{Computational Behavior}

Figs.~1 and 2 demonstrate that each new cycle of partial peeling of the tape forms a memory and, in the process, may erase an existing memory, depending on the sequence of the inputs. However, the most recently encoded value is always the first to be read out. This feature allows us to demonstrate a simple computing behavior that compares successive inputs. In Fig.~2(c), the nature of $F_p$ during the encoding of the second input in both sequences (highlighted by the pink dashed region) represents a decision: if the present input/turning point is smaller than the previous one, the peeling force is monotonic; if the present input is larger, the force shows a spike that quantifies the previous value. This purely mechanical computation resembles one-back comparison tasks used to study working memory in neuroscience.

\begin{figure}
 \centering
        \includegraphics[width=0.9\textwidth]{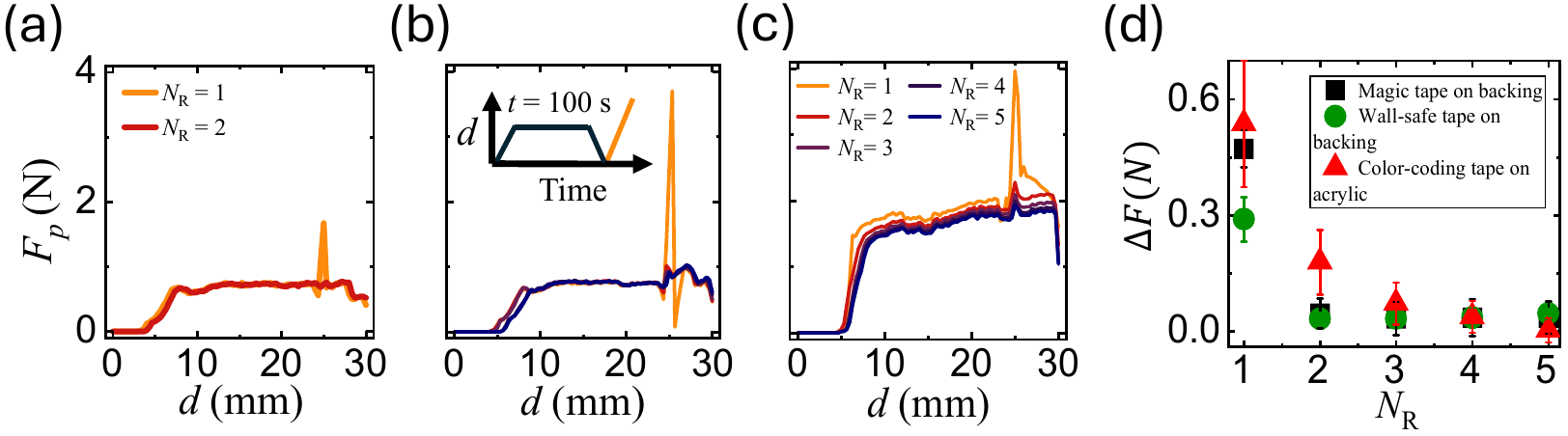}
 \caption{\textbf{Tuning memory strength and partial erasure.} (a)--(c) Peeling force plotted during readouts for an encoded memory of $d=25~mm$. $N_R$ indicates number of repeated readout sweeps. 
 (a) Peeling force during two consecutive readouts. The second readout shows almost no signature at $d=25~mm$, implying complete erasure by the first readout. 
 (b) Readout after 100~s aging (schematic of protocol in inset). The memory is much stronger in the first readout, and there is a vestige in every subsequent readout. 
 (c) Readout of tape on a smooth acrylic substrate. Even without significant aging time, the memory persists. 
 (d) Relative memory strength $\Delta F = F_p(\mathrm{peak}) - F_p(\mathrm{baseline})$ in successive readouts. Black squares are from panel (a), and green triangles from panel (c) (see Materials and Methods). 
$F_p(\mathrm{baseline})$ is the value of $F_p$ estimated at $d$ in the absence of a peak, using a linear fit of the underlying peeling force. Error bars are estimated from three different samples of each tape. 
 }
\label{fig3}
\end{figure}

\subsection{Tunable memory strength and partial erasure}

The previous experiments showed that a single cycle of encoding forms a memory with a signature that approximately doubles the peeling force and that peeling past a stored memory erases it. These findings are summarized in Fig.~3(a), which shows the results of two successive readout cycles. We now test whether either or both of these behaviors can be tuned. At the outset, we note that reinforcing a memory with an additional cycle of encoding does not appreciably strengthen the memory signature. However, the unique physics of tape offers other ways to tune the memory signature and erasure.

We first amplify the memory signature by adding a pause at the turning point of $d=25$~mm for 100~s (schematic shown in the inset of Fig.~3(b)), before laying the tape back down and reading out the memory. Such an ``aging'' protocol applies prolonged stress at the peeling front, locally strengthening the pressure-sensitive bond~\cite{Bartlett2023Aug, Villey2015Apr}. %
The resulting memory signature is much stronger compared to Fig.~3(a), as indicated by the enhanced spike in panel (b). Further, this stronger memory is also persistent: subsequent readouts ($N_R =2$ to 5) show a residual signature of the memory rather than the prompt erasure observed so far. We note that even without a pause in the protocol, rapid initial aging may be the reason for the memory-indicating spike in the force that is so much greater than the baseline force for steady peeling. We also find that a variation of the aging protocol can \emph{weaken} the encoded memory; instead of being held taut, if the tape is left flat on the substrate (after writing the memory) for an extended time, the baseline force grows stronger while the peak force that signifies the memory saturates. The perceived memory signature therefore becomes less distinct. 

Fig.~3(b) indicates that a stronger memory may erase more gradually. To confirm this, we can instead enhance the adhesion strength by keeping the same protocol as Fig.~3(a) but changing the materials. We change the substrate from tape backing to smooth cast acrylic (see Materials and Methods) in Fig.~3(c). The tape inherently has a stronger adhesion to the acrylic substrate as indicated by the higher baseline peeling force ($\sim 2$~N in Fig.~3(c), whereas in Figs.~3(a) and 3(b), it is $\sim 1$~N). Despite this, the spike in $F_p$ is still significantly stronger than in Fig.~3(a) and again, we see clear residual memory signatures that persist over several readouts. 

The tunability of memory strength in tape thus offers a rare opportunity to vary erasure. In Fig.~3(d), we summarize the erasure behavior of other tape-substrate combinations (see caption and Materials and Methods for details). To incorporate differences in the overall adhesion strengths for the various tape-substrate combinations, we quantify the relative memory strength $\Delta F = F_p (\mathrm{peak}) - F_p (\mathrm{baseline})$ for subsequent readout cycles $N_R$. Notably, color-coding tape shows memory that grows weaker over multiple cycles.

\section{Latching model}
\begin{figure}
 \centering
        \includegraphics[width=0.9\textwidth]{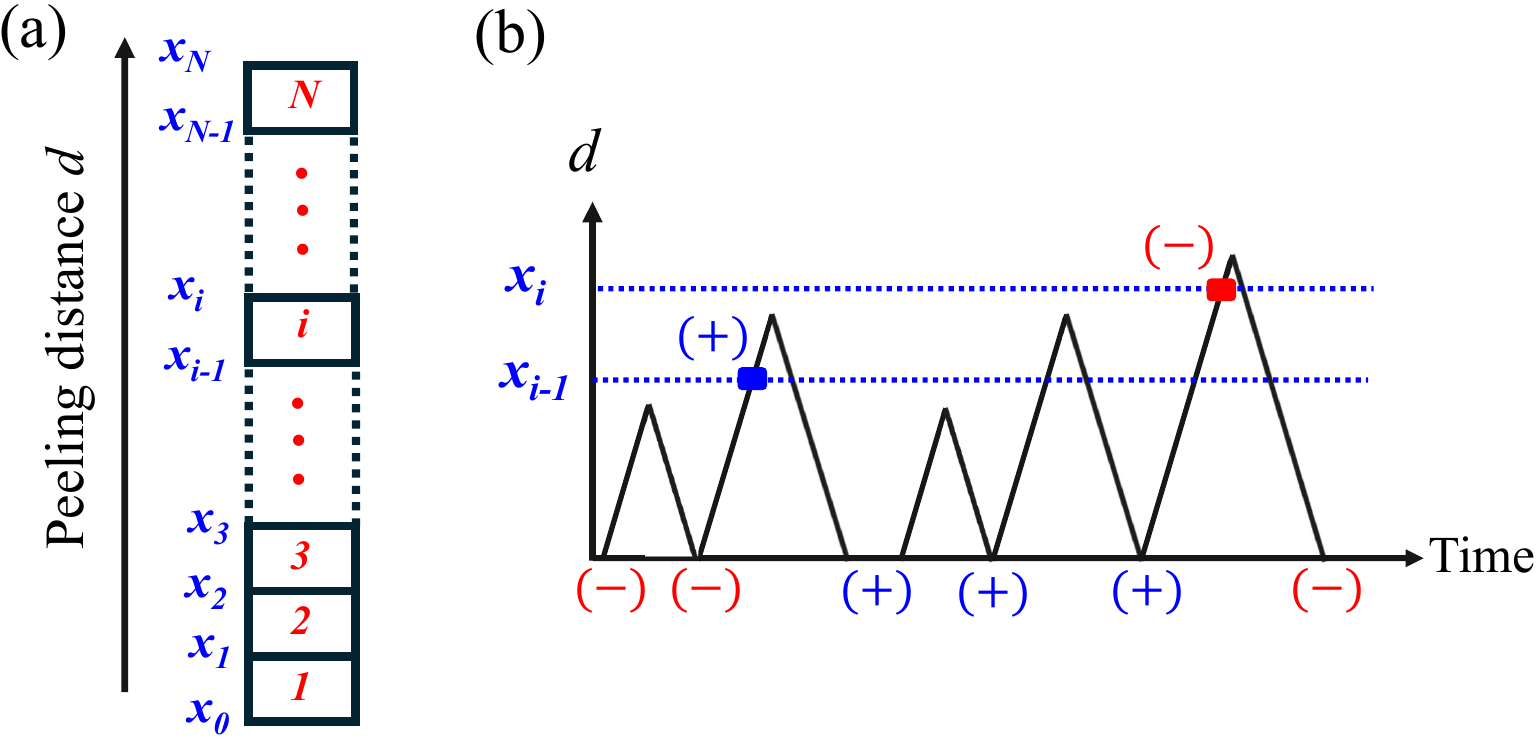}
 \caption{\textbf{Minimal model for the tape's memory behavior.}
 (a) The tape represented as an array of bits, with boundaries at the $x$ values below and above each bit. %
 (b) Behavior of a bit under external driving. Each bit starts out in the ($-$) state. When the peeling distance $d$ crosses the lower threshold $x_{i-1}$, the bit flips to a ($+$) state, corresponding to strong adhesion. It remains ``latched'' in this state until $d$ crosses the upper threshold $x_i$ and resets the bit to ($-$).} 
\label{fig4}
\end{figure}

For the case of complete erasure (Fig. 2), the global memory behavior of the tape arises from the local continuum physics of adhesion and elasticity. We now seek a much simpler representation of the tape as an array of independent mesoscopic bits. 
The bits in Fig.~4(a) represent segments of the tape, with the $i$th bit having lower and upper boundaries $x_{i-1}$ and $x_i$. 
Fig.~4(b)shows the functioning of each bit. The bit starts in a ($-$) state, corresponding to the weak adhesion. When the external perturbation crosses the lower threshold of the bit ($x_{i-1}$), the bit flips to a ($+$) state, corresponding to strong adhesion. It remains in this state until the perturbation crosses the upper threshold of the bit ($x_i$) which resets it back to the ($-$) state. 

This set of rules resembles the ``latching'' mechanism proposed by Lindeman et al.~\cite{Lindeman2025Jan}. While that work focused on interacting pairs of bistable elements (``hysterons''), in both instances, we see a subsystem that becomes stuck in a new state when driven, with the only possibility of a reset being a larger drive. In that theoretical work and a subsequent experimental demonstration by Paulsen~\cite{Paulsen2024Sep}, it was envisioned that a system could efficiently store multiple memories if its latches' thresholds were fine-tuned according to the scheme of Fig.~4(a). In adhesive tape, this scheme involves no tuning because the latches' thresholds are set by their spatial arrangement. Memory capacity is also straightforward: the number of bits $N$ scales linearly with tape length, and the state of the tape distinguishes among $2^N$ different sequences of peeling.

\section{Discussion}

An everyday object---pressure-sensitive adhesive tape---let us explore multiple memories of rectified mechanical perturbations. We studied mechanical memory retention and loss to establish the generic rules that govern the memory behavior of the tape. We also showed how this simple memory leads to function, demonstrating a mechanical comparator with parallels to cognitive tasks involving working memory. These findings encourage further exploration of neuroscience-inspired memory behavior in physical systems.

The broader significance of our results for material memory comes from the comparison with return-point memory. Return-point memory is the generic way for a system to promptly form and recall memories of nested turning points, yet the behavior cannot store multiple memories of rectified driving. 
It was originally noted in ferromagnets, and it was formally demonstrated in a disordered ensemble of hysterons, which is a simple model of magnetic hysteresis~\cite{Adair1983Apr, Preisach1935May}. However, the behavior is also reliably obtained in systems whose elements have crucial differences from hysterons. Two recent examples are amorphous solids, where these elements can interact strongly, can be easily created or destroyed, have dynamical behaviors, and may sometimes have more than two states~\cite{Paulsen2025Mar, Lindeman2025Jan}; and the single-dial combination lock~\cite{Keim2019Jul}, with continuous degrees of freedom and no disorder.
Thus, a wider variety of mesoscopic physics can give rise to the same global property~\cite{Sethna1993May, Mungan2019May}.
Our present work takes a major step toward an analogous general principle for promptly storing and erasing nested memories of rectified driving, by showing how the peeling of adhesive tape---an intricate problem that involves rheology and elasticity on multiple scales---gives rise to the same global behavior as interacting pairs of hysterons.

In considering these computational and theoretical aspects, we have focused on complete, prompt erasure. However, our experiments also showed that increasing the adhesion strength can make a memory persist over many cycles, when it would otherwise be erased. This partial erasure does not fit neatly into the latching framework, and would suggest a different computational function. We found that enhanced adhesion can be tuned either by adjusting the protocol (aging), or by changing materials to increase affinity (tack). This convenient tunability stands in sharp contrast with other systems. The mechanism for persistent memories is an open question, and could include plastic deformation of the backing, or a change in the adhesive layer itself.

While our work establishes tape as an appealing platform for tunable mechanical memory, pressure-sensitive tape has broad commercial and technological value, and peeling of tapes has long served as a model system for diverse problems in fracture mechanics and adhesion physics. However, cyclic, partial peeling protocols have received little attention to date. The unusual memory phenomena in our work suggest an area for fundamental research at the intersection of thin-sheet mechanics and adhesion.

\section{Materials and Methods}
Unless otherwise noted, our experiments use 3M ``Scotch Magic'' tape of width 12.7~mm (1/2 inch). A layer of Magic tape firmly applied to a plate of cast acrylic (PMMA) is used as the fixed substrate. In Fig.~3, we also use 3M ``Scotch Wall-Safe'' tape of width 25.4~mm (1 inch) and red 3M ``Scotch 690 Color Coding'' tape of width 6.35~mm (1/4 inch). The substrate for the wall-safe tape is a piece of the same tape adhered to the acrylic surface. Data on color-coding tape (Fig.~3(d)) are taken with the acrylic surface as the substrate.

The experimental setup consists of the tape placed on the substrate, with the peeling end connected to the shaft of a motorized Baoshishan ZP-500N force gauge mounted to a vertically oriented Newport SMC100 translation stage. Since we control the stage's movement, all distances mentioned in the text correspond to movement of the stage to a distance $d$. To encode and read memories in the tape, the stage and the force gauge move up to a distance $d$ and down to the initial position at $d=0$. The stage moves with a speed of 3~mm/s throughout the process, and the reversal includes a pause of $\sim1$~s. %
The up and down motion causes the tape to peel off and lie back down. The translation stage motion is programmed by a Python script. Images from a camera mounted $\sim$0.5~m above the substrate are used to correlate the peeled distance with the force readouts.

\ack{We thank Omri Barak, Matthew Diamond, Anupam Pandey, and Ye Xu for helpful discussions. }

\funding{This work was supported by Research Grant RGP0017/2021 from the Human Frontier Science Program. CCM was supported by the Penn State Department of Physics and the Center for Nanoscale Science (NSF-MRSEC) and the National Science Foundation (DMR 2011839 and PHYS 2349159).
}

\roles{CCM built the force measurement setup and performed and analyzed most of the experiments. SC performed additional experiments and supervised. SC and NK conceptualized the research and wrote the paper, with help from CCM.}

\printbibliography

\end{document}